\documentclass[reqno]{amsart}
\usepackage{graphicx}
\usepackage{gensymb}                   
\usepackage{url}
\theoremstyle{definition}

\theoremstyle{remark}

\begin{document}

\title[Henry's Law]{No reliable studies of climate change without Henry's Law and a new thermometer for the global temperature}%
\author{Jyrki Kauppinen and Pekka Malmi}%
\address{The authors are retired from the Department of Physics and Astronomy, University of Turku, Finland.}%
\email{jyrki.kauppinen@utu.fi and jyrki.kauppinen@gasera.fi}%


\date{\today}%
\begin{abstract}

In our previous paper ``No experimental evidence for the significant anthropogenic climate change'' we had a reference to this paper. Thus, we have presented a new theory: how Henry's Law regulates the concentration of CO$_2$ in the atmosphere. This theory uses a physically perfect unit impulse response to convolve signals. By comparing the theory and the present observations we are able to derive the response time about 7.4 years, which is roughly the residence time of CO$_2$ in the atmosphere. In addition, according to Henry's Law we can derive the temperature dependence of the equilibrium concentration of CO$_2$. It turns out that now the major increase of CO$_2$ in the atmosphere is due to the temperature change, the reason for which is approximately a change of cloudiness ($90\%$) and the rest ($10\%$) the greenhouse effect. The corresponding increase of CO$_2$ is $83$~ppm/$\degree$C. According to our theory, the causality is very clear: the increase of the temperature results in more CO$_2$ in the atmosphere. This further CO$_2$ concentration increases also the temperature but less than $10\%$ via the greenhouse effect. The human contribution to the CO$_2$ concentration and the temperature is very small. About $90\%$ of a big human release of CO$_2$ dissolves in water. That is why the human release now warms the climate less than $0.03\degree$C. We can also perform an inverse operation or calculate the temperature profile from the observed CO$_2$ curve. So we propose a new thermometer for the global surface temperature. This can be used almost in real time and it is more accurate than the present methods. It turns out that probably also N$_2$O and SF$_6$ concentrations in the atmosphere operate as global thermometers.
\end{abstract}
\maketitle
\section{Introduction}

The study of climate change requires knowledge of the gas components in the atmosphere. In addition to water vapor we have to know other components like CO$_2$, CH$_4$, and N$_2$O. Because more than $71\%$ of the surface of Earth is water, we have to be aware of the gas exchange between water and the atmosphere. This is possible with the help of Henry's Law, which states that the solubility of a gas in a liquid is proportional to the partial pressure of the gas above the liquid. The proportional coefficient or the Henry's Law constant depends on the liquid-gas pair and the temperature. So we have to conclude that all the papers and studies without the use of Henry's Law cannot give the reliable results of climate change. This is true in the IPCC reports. Many books like ``Global Physical Climatology'' \cite{Hartmann} dealing with climate change do not even mention Henry's Law. Because the Henry's Law constant depends on the temperature, concentrations of gases emitted from water give the global temperature behavior. It turned out that our method results in even more accurate global temperatures than the used methods. That is why we propose a new thermometer of the global temperature.

\section{Application of Henry's Law to the Carbon cycle}

As mentioned before Henry's Law \cite{Henry1803,enwiki:1136260189,Harvey2007,Sander2015} is
\begin{equation}\label{H}
    c_l=pH,
\end{equation}
where $c_l$ is a concentration of a dissolved gas in liquid (mol/L), $p$ is a partial pressure of the gas in the atmosphere (atm), and $H$ is the Henry's Law constant $3.4\cdot 10^{-2}$ mol/(L atm) for CO$_2$, when $T=289.15$~K. Equation~(\ref{H}) is valid only in an equilibrium. Nature always changes $c_l$ and $p$ towards the equilibrium with a certain response time $\tau$. In the century between $1700$-$1800$, the atmosphere and water were in the equilibrium and $T$ was $T_0$ and the concentration of CO$_2$ was $280$~ppm. $T_0$ was probably about $287$~K or $14\degree$C. When the temperature $T$ started to increase from $T_0$, then $H$ decreased and $p$ increased slowly from $p_0$ ($280$~ppm) so that Eq.~(\ref{H}) was approximately valid. Later we will use for $p$ unit ppm ($7.84$~Gtn) and $\Delta T(t)=T(t)-T_0$. We define a new parameter or the global emission strength $\alpha=dp/dT$~(ppm/$\degree$C), which gives a new equilibrium concentration $280\,{\rm ppm}+\alpha\Delta T(t)$. When climate is out of the equilibrium we have to know how it goes to the equilibrium as a function of time. The unit step response is the solution of the following differential equation: $dp/dt=(1-p)/\tau$. This solution is $p=1-\exp(-t/\tau)$, whose derivative is the unit impulse response $I(t)=(1/\tau)\exp(-t/\tau)$, if $t>0$, and $0$, if $t<0$ \cite{Kau2001,Kau2011}. This means that the observed concentration is given by the convolution of $I(t)$ and $p(t)$ as follows
\begin{equation}\label{convolution}
    I(t)\ast p(t)=\int_{-\infty}^{\infty}I(u)p(t-u)du.
\end{equation}
Because the net CO$_2$ dissolved in water is $I(t)\ast (p_r(t)-\alpha\Delta T(t))$, then in the atmosphere the measured CO$_2$ concentration $p_m(t)$ has to be as follows
\begin{align}\label{pm}
    p_m(t) &= p_r(t)-I(t)\ast(p_r(t)-\alpha\Delta T(t))+280~{\rm ppm} \\
           &= p_r(t)-I(t)\ast p_r(t)+\alpha I(t)\ast\Delta T(t)+280~{\rm ppm} \nonumber \\
           &= p_h(t)+\alpha I(t)\ast\Delta T(t)+280~{\rm ppm}. \nonumber
\end{align}
In the above equation $p_r(t)$ is the human release of CO$_2$ to the atmosphere and $p_h(t)$ is the part of the human release, which stays in the atmosphere. More of the properties of convolution are presented in our book \cite{Kau2001}. Next we have to calculate the human contributions in Eqs.~(\ref{pm}). We are lucky that Hermann Harde \cite{Harde2019,LeQuere,FossilFuel2017}, has fitted in an exponential form the estimated anthropogenic CO$_2$ release rate $dp_r(t)/dt$, which includes the land use, too. In the fit the time interval was between 1850 and 2010. The result is given by
\begin{equation}\label{pr'}
    \frac{dp_r(t)}{dt}=A\left(\exp\left(\frac{t-t_0}{\tau_e}\right)+4\right),
\end{equation}
with the parameters: $A=0.026$~ppm/yr, $t_0=1750$~yr, and $\tau_e=50$~yr. The integral over the fitted release rate $dp_r(t)/dt$ above gives the total cumulative release in ppm
\begin{equation}\label{pr}
    p_r(t)=1.3\,\exp\left(\frac{t-t_0}{\tau_e}\right)+0.104\,(t-t_0).
\end{equation}
This agrees very well with the observation. Note that $p_r(t)$ goes directly to the atmosphere. The advantage of the above analytical form of $p_r(t)$ is that we can calculate an analytical expression of the convolution $I(t)\ast p_r(t)$, because $p_r(t)$ has exponential and linear terms. So we can calculate $p_r(t)-I(t)\ast p_r(t)=p_h(t)$ or the anthropogenic concentration, which stays in the atmosphere from the release $p_r(t)$. Note that $I(t)\ast p_r(t)$ dissolves in water. Thus the result is given by
\begin{equation}\label{ph}
    p_h(t)=\frac{1.3\,\tau}{\tau+\tau_e}\exp\left(\frac{t-t_0}{\tau_e}\right)+0.104\,\tau.
\end{equation}
If $dp_r(t)/dt$ is a constant, then $p_h(t)=\tau dp_r(t)/dt$, which is useful in the estimation of carbon sinks.
In order to attain our final goal, i.e., to derive $\alpha$ and $\tau$ we have two possibilities. First we calculate the convolution $\alpha I(t)\ast\Delta T(t)=p_e(t)$. Now according to Eqs.~(\ref{pm}) $p_h(t)+p_e(t)+280~{\rm ppm}$ should be the measured CO$_2$ concentration $p_m(t)$ (the smoothed Keeling curve) in the atmosphere. Thus, by comparing the calculated concentration with the measured one $p_m(t)$ we are able to derive $\tau$ and $\alpha$. However, the computations of convolution integrals are a little problematic, because in convolution at a certain point requires measured data even $3\tau$ earlier than this point. Convolution has a memory effect. The second possibility is to carry out the calculation backwards, i.e., the calculation of $\Delta T(t)$ via the deconvolution of $p_e(t)$ by $\alpha I(t)$. This can be done using the Fourier transforms \cite{Kau2001}, but that is quite complicated. It is better to derive a new equation for the deconvolution by $(\alpha/\tau)\exp(-t/\tau)$. The derivative $dp_e(t)/dt=\alpha dI(t)/dt\ast\Delta T(t)=\alpha(\delta(t)/\tau-\exp(-t/\tau)/\tau^2)\ast \Delta T(t)=(\alpha\Delta T(t)-p_e(t))/\tau$, because $\delta(t)$ is Dirac delta function. Thus, we have
\begin{equation}\label{DT}
    \Delta T(t)=\frac{1}{\alpha}\left(p_e(t)+\tau \frac{dp_e(t)}{dt}\right),
\end{equation}
where $p_e(t)=p_m(t)-280\,{\rm ppm}-p_h(t)$ according to Eqs.~(\ref{pm}). Using the above equation we need only derivation instead of integration or the Fourier transforms. Thus, by comparing calculated $\Delta T(t)$ with the measured one $\Delta T_m(t)$ we are able to fit $\tau$ and $\alpha$. This method is very powerful to measure a global temperature in the absolute scale starting from $T_0$. Thus, we propose it for use as a thermometer for the global temperature.

\begin{figure}[h]
  \centering
  \includegraphics[width=\columnwidth]{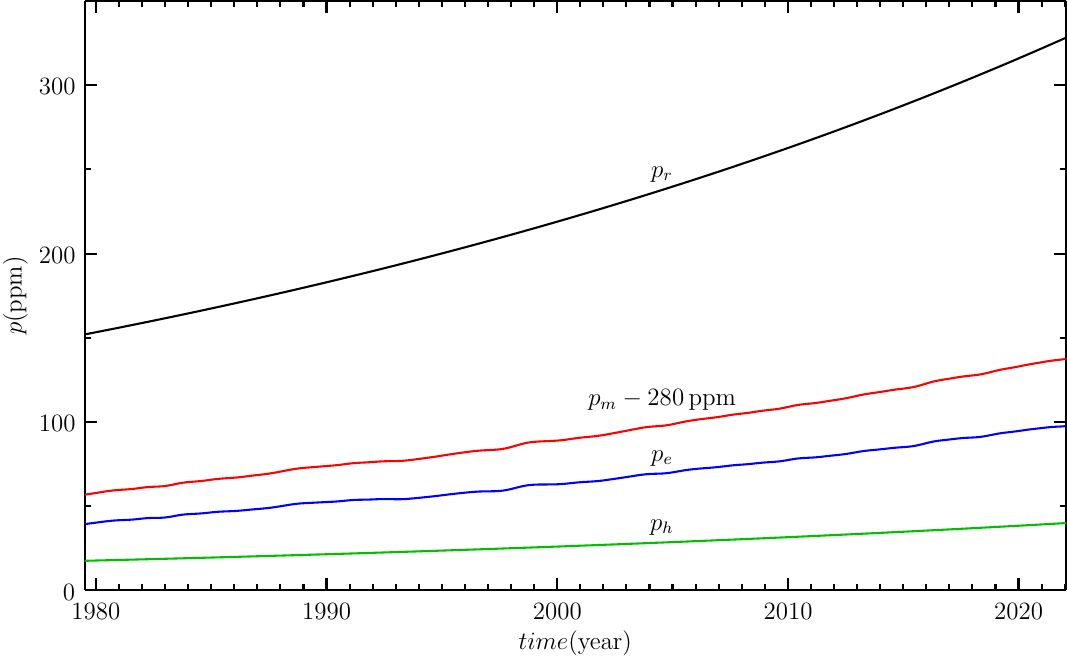}
  \caption{\footnotesize Human CO$_2$ release $p_r(t)$ (black), measured $p_m(t)-280\,$ppm (red), CO$_2$ emitted by water $p_e(t)$ (blue) and human CO$_2$ contribution to the atmosphere $p_h(t)$ (green).}\label{p}
\end{figure}

Figure~\ref{p} shows the functions needed in computation: Human cumulative released $p_r(t)$ Eq.~(\ref{pr}), $p_m(t)-280\,{\rm ppm}$ i.e., about one year Fourier-smoothed \cite{Kau2001,Kau1982} Keeling curve, $p_h(t)$ Eq.~(\ref{ph}) and finally $p_e(t)=p_m(t)-280\,{\rm ppm}-p_h(t)$. Note that the human contribution $p_h(t)$ is small only about $12\%$ of $p_r(t)$.

\section{Experimental derivation of the response time $\tau$ and the global emission strength $\alpha$}

\begin{figure}
  \centering
  \includegraphics[width=\linewidth]{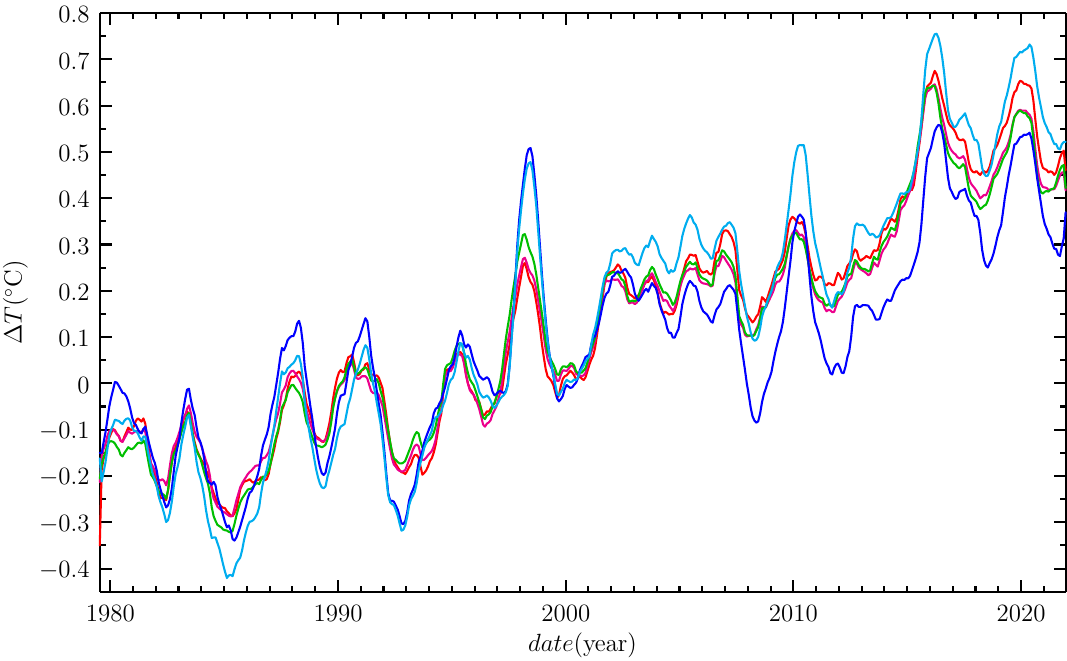}
  \caption{\footnotesize One year running averages of global monthly temperature anomalies normalised by comparing to the average value of 30 years from January 1979 to December 2008 \cite{Humlum}. UAH MSU (blue), RSS MSU (cyan), GISS (red), NCDC (magenta) and HadCRUT5 (green).}\label{HumlumTemperatures}
\end{figure}

Figure~\ref{HumlumTemperatures} shows five observed temperature anomalies $\Delta T_a(t)$ according to Ole Humlum \cite{Humlum}. As you can recognize, these anomalies deviate quite a lot from each other, even $0.425\degree$C. However, the overall temperature increase is clear and El Niño, too. The anomalies UAH MSU and RSS MSU oscillate more than the three others. Even those better three anomalies differ of the order of $0.15\degree$C from each other.

\begin{figure}[h]
  \centering
  \includegraphics[width=\linewidth]{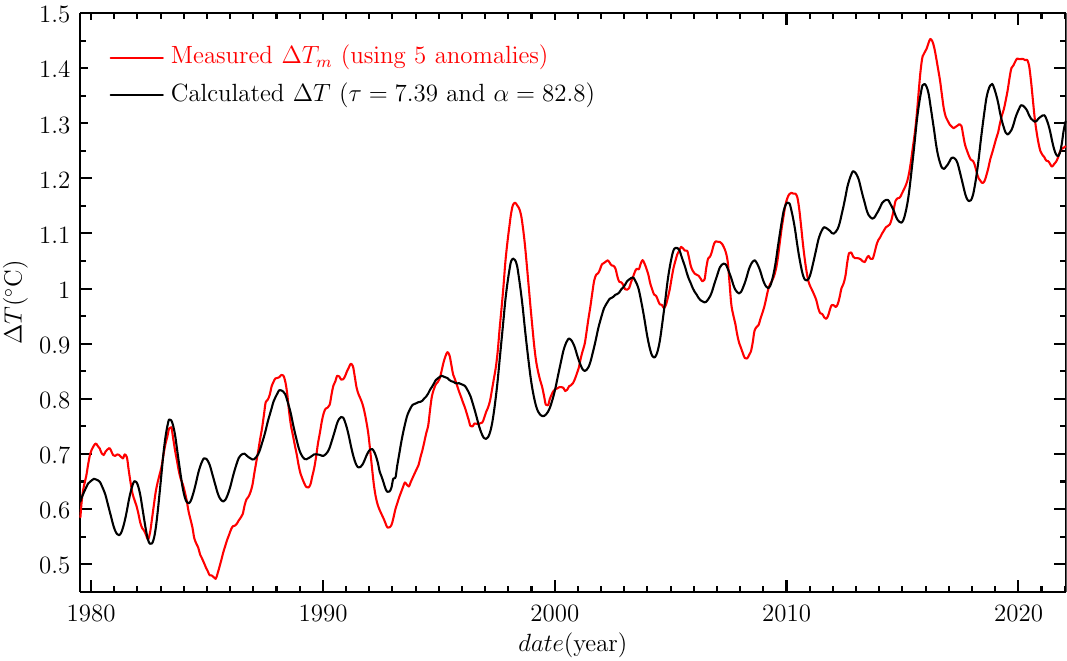}
  \caption{\footnotesize The red curve is the average of all the five anomalies in Fig.~\ref{HumlumTemperatures} added by $\Delta T_{ref}=0.801\degree$C. The black curve is calculated from Eq.~(\ref{DT}) with $\alpha=82.8\,{\rm ppm}/\degree{\rm C}$ and $\tau=7.39\,{\rm yr}$.}\label{TmTc}
\end{figure}

Now we are ready to derive $\alpha$ and $\tau$ so that the calculated temperature $\Delta T(t)$ according to Eq.~(\ref{DT}) explains the measured temperature $\Delta T_m(t)$, which is the average of the anomalies $\Delta T_a(t)+\Delta T_{ref}$. We carry out a nonlinear least squares fit of the function $\Delta T(t)$ to $\Delta T_m(t)$. The extra fitting parameter $\Delta T_{ref}$ is the temperature, which we have to add to the anomalies so that they reach the level of the calculated $\Delta T(t)$. Using the average of all the five anomalies smoothed by one year moving average in $\Delta T_m(t)$ the fit gave $\alpha=82.7\,{\rm ppm}/\degree$C, $\tau=7.43$~yr and $\Delta T_{ref}=0.801\degree$C. This fit is the calibration of the thermometer. Most of the reported residence times are $5-10$ years \cite{Solomon}. The final results are seen next in Fig.~\ref{TmTc}. We repeated the above fit using only three better measured temperature anomalies and their average. This fit resulted in $\alpha=84.9\,{\rm ppm}/\degree$C and $\tau=6.30$~yr.

\begin{figure}
  \centering
  \includegraphics[width=\linewidth]{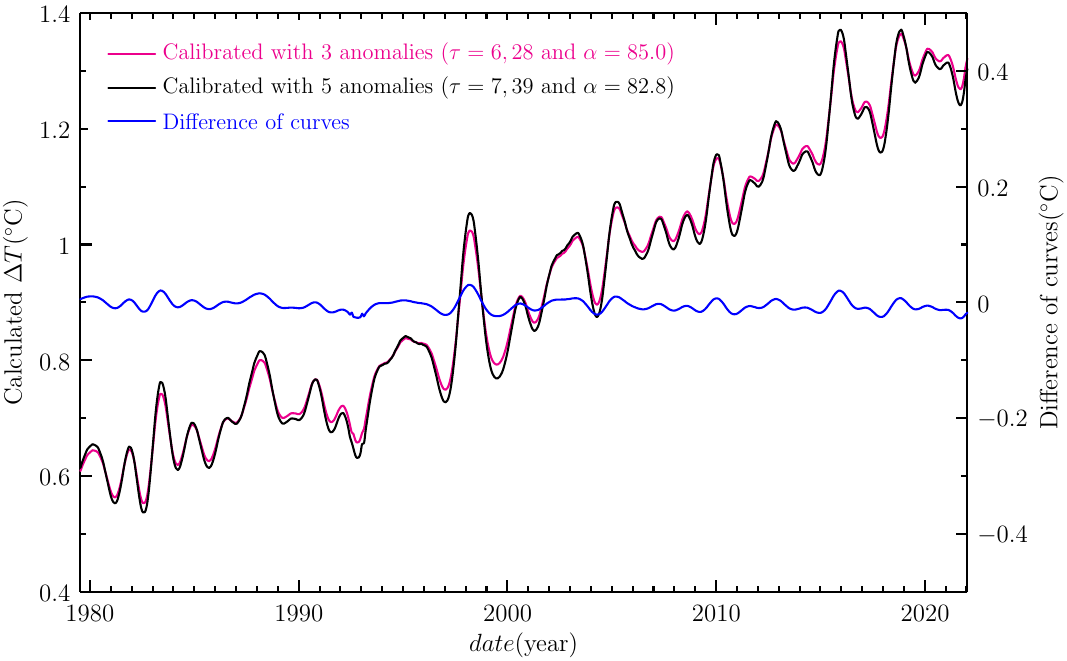}
  \caption{\footnotesize The temperatures calculated using the average of the temperature anomalies GISS, NCDC and HadCRUT5 with  $\tau=6.28\,{\rm yr}$ and $\alpha=85.0\,{\rm ppm}/\degree{\rm C}$ (purple) and the temperature anomalies GISS, NCDC, HadCRUT5, UAHMSU and RSSMSU with $\tau=7.39\,{\rm yr}$ and $\alpha=82.8\,{\rm ppm}/\degree{\rm C}$ (black).}\label{Tc3Tc5}
\end{figure}

\begin{figure}
  \centering
  \includegraphics[width=\linewidth]{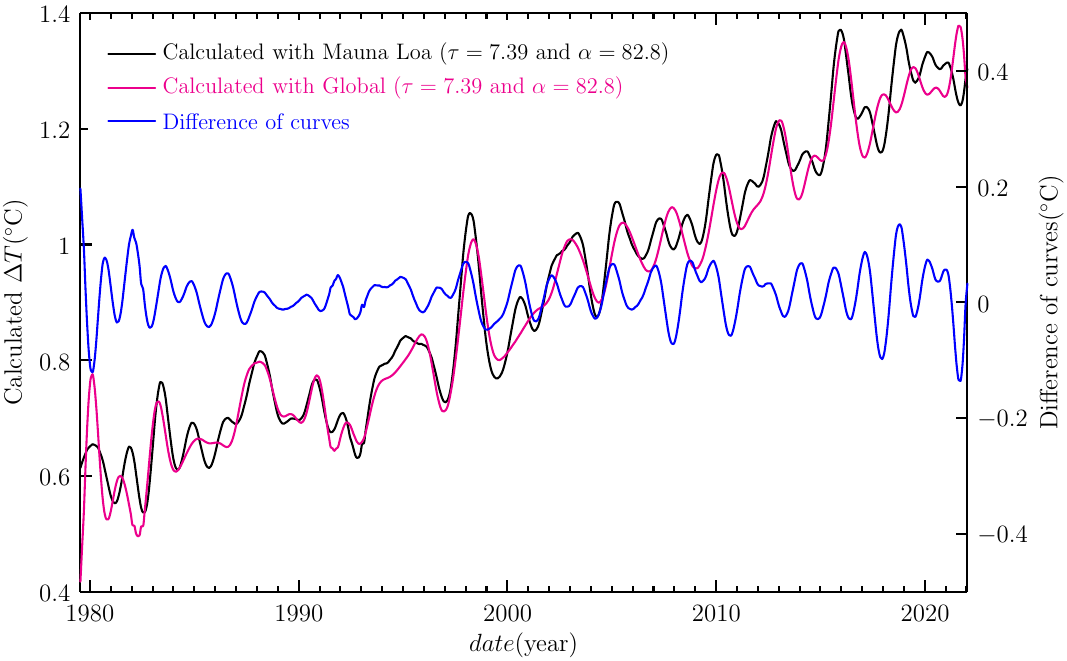}
  \caption{\footnotesize The temperatures calculated using the average of the temperature anomalies GISS, NCDC, HadCRUT5, UAHMSU, RSSMSU and the Mauna Loa CO$_2$ curve (black) and the global CO$_2$ curve (magenta).}\label{TMLTGL}
\end{figure}

Figure~\ref{Tc3Tc5} shows the temperatures calculated using the above derived $\alpha$ and $\tau$ in the cases, where we used three and five measured temperature anomalies. This figure gives us a hint how much the used measured temperature anomalies affect the precision of the thermometer via the parameters $\alpha$ and $\tau$. Even though the measured temperature anomalies are quite inaccurate, the maximum error of the calculated temperature is only $0.03\degree$C.
In figure \ref{TMLTGL} we have tested the accuracy of the method caused by possible errors in the measured $p_m(t)$.
Two temperature curves are calculated using as $p_m(t)$ the Mauna Loa \cite{MaunaLoa} and the global CO$_2$ \cite{globalCO2} curves. As you can see, differences are mainly well below $0.1\degree$C. However, the errors of $p_m(t)$ dominate. Thus, according to Figures \ref{Tc3Tc5} and \ref{TMLTGL} we can conclude that the precision of our global thermometer is better than $0.1\degree$C using only a single measured CO$_2$ curve. Of course, the accuracy can be improved using the average of many CO$_2$ observations and a longer smoothing time.

\section{Causality: Temperature is a cause of CO$_2$}

\begin{figure}[t]
  \centering
  \includegraphics[width=\linewidth]{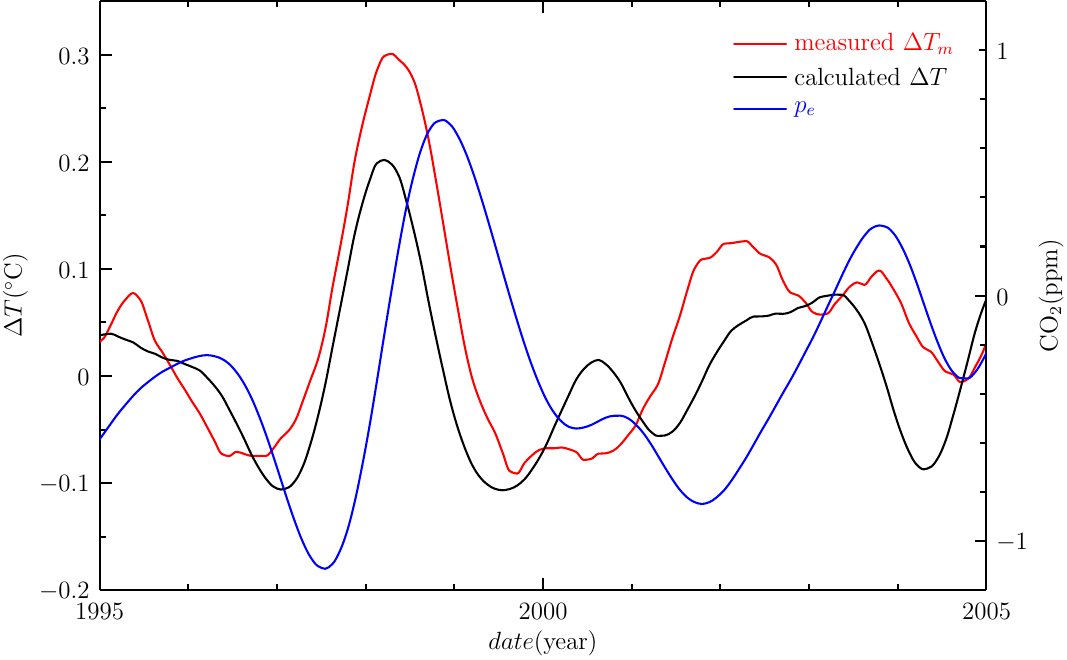}
  \caption{\footnotesize A detail of Fig.~\ref{TmTc} around the El Niño peak 1998 superimposed with the CO$_2$ curve $p_e$ (blue).}\label{ElNino}
\end{figure}

According to our theory, the causality means that the temperature change $\Delta T(t)$ produces the most CO$_2$ via the convolution $p_e(t)=\alpha I(t)\ast\Delta T(t)$. It is well known that effects never precede their causes. The causality is the basic property of the convolution by $I(t)$, which is zero when $t<0$. With the help of Fig.~\ref{ElNino} we will demonstrate the causality using the measured signals in a real case.

Figure~\ref{ElNino} shows the measured temperature $\Delta T_m(t)$, the measured CO$_2$ concentration $p_e(t)$, and the calculated temperature $\Delta T(t)$ around the El Niño peak 1998. We have subtracted from the above curves a parabolic baseline in order to see better the El Niño peaks. In this figure you can find that the measured $\Delta T_m(t)$ and the calculated $\Delta T(t)$ are very similar, but the measured CO$_2$ in $p_e(t)$ is delayed about one quarter its period. This is the proof that the temperature change causes the major CO$_2$ change, not vice versa. Ole Humlum has studied experimentally the same El Niño peak and its delay \cite{Humlum2013}. Our results are in a very good agreement with those of Ole Humlum.

According to IPCC, the main cause of climate change is CO$_2$. Thus, using the climate sensitivity we can calculate the temperature peak from El Niño CO$_2$ peak of about $1.8$~ppm in Fig.~\ref{ElNino}. So, according to IPCC, the amplitude of the temperature peak should be about $0.24\degree{\rm C}\,\ln((420+1.8)/420)/\ln2=0.00148\degree$C. In addition, the peak has the further delay of about one month. This should be the amplitude of the observed or the calculated temperature peak in Fig.~\ref{ElNino}. However, the measured peak is about $0.37\degree$C and the calculated one $0.32\degree$C. The calculated peak is approximately $\tau/\alpha$ times the derivative of $p_e(t)$. All strongest real temperature peaks are followed by lagged CO$_2$ peaks with a similar manner as the El Niño peak. See also the figures of Ole Humlum \cite{Humlum2013}. The above calculation means that the method of IPCC to calculate temperature changes is completely wrong in magnitude, even if we use ten times higher sensitivity like IPCC. The most important thing is the fact that the IPCC method produces the temperature peak at a wrong position nearly one year later than the measured one. These are the strong proofs, that the causality used in IPCC reports is wrong. Thus, we have proven that merely the greenhouse effect cannot explain climate change.

\section{Carbon sinks}

If the human release $p_r(t)$ and $p_h(t)$ go rapidly to zero, then after $3\tau$ the global temperature has dropped about $0.03\degree$C. This gives us a hint, how much we can affect the temperature using carbon sinks. The forests are good carbon sinks. We have land less than $29\%$ of the area of Earth. About one third of land is coved by forests, which absorbs CO$_2$ about 15~Gtn/yr. In principle, we can double the area of forests, which lowers the temperature approximately $0.01\degree$C.

\section{Discussion}

One very important advantage of the calculated $\Delta T(t)$ is that it gives the temperature in the right and real scale, whose zero value is on the level, where CO$_2$ is $280$~ppm.

As shown in Fig.~\ref{Tc3Tc5}, there is a clear need of the more accurate measurement of the global temperature. Carbon dioxide is an indicator of the global temperature. That is why we have proposed a new thermometer for the global temperature based on the Eq.~(\ref{DT}). We can use every month measured CO$_2$ data $p_m(t)$ at Mauna Loa to calculate the points in $p_e(t)$ by subtracting the known human release $p_h(t)$ from $p_m(t)$. So, we can calculate a new temperature value every month but six months later, if we smooth $p_m(t)$ by one year moving average.

It is important to realize that Henry's Law uses all the possible water surfaces: oceans, lakes, rivers, and even water drops in rainfall and on the leaves of trees. In addition, in the atmosphere all the gases obey Henry's Law. However, we do not need to know in detail the ocean chemistry and the Henry's Law constant, because we use measured global values of $p_m(t)$, $p_r(t)$, and $\Delta T_m(t)$. These values included all the needed knowledge. In addition we do not like to handle the problem locally like James R. Barrante \cite{Bar2014}, but instead we use global values. The net CO$_2$ dissolved in water mainly in the oceans is $I(t)\ast(p_r(t)-\alpha\Delta T(t))$, which acidifies the oceans so that pH is decreasing. The calculated temperature $\Delta T$ roughly consists of $-11\degree{\rm C}\,\Delta c+{\rm GH+GH(human)}$, where $\Delta c$ is a change of low cloud cover, GH is a greenhouse term of the extra CO$_2$ due to the warming of $-11\degree{\rm C}\,\Delta c$, and ${\rm GH(human)}=0.24\degree{\rm C}\,\ln((420+p_h(t))/420)/\ln2=0.029\degree{\rm C}$ in 2020. At last, we have to point out that $\tau$ and $\alpha$ derived in this paper are valid in a limited temperature range. Theoretically $\alpha$ depends on the temperature slightly exponentially. The study of this question requires more precise measured temperatures. On the other hand we have the Keeling curve down to 1960 so we can calculate the temperature down to 1960, too. Around 1973 it was a quite strong El Niño. We do not show this part of the temperature, because there is no good measured temperature for comparison.

\section{Conclusion}

In an equilibrium, Henry's Law is very trivial and we do not need to calculate any convolutions. But, if the climate is out of the equilibrium, then we have to apply the unit impulse response via convolution in order to describe concentrations as a function of time. Right now, our climate is substantially out of balance, because of the big human CO$_2$ release and the change of low cloud cover. In this situation, we must use our theory to study climate change.

Nature convolves the true temperature $\Delta T(t)$ by $\alpha I(t)$. The convolution is $p_e(t)$ the CO$_2$ concentration emitted to the atmosphere. According to Eq.~(\ref{DT}), deconvolution back is very simple $\Delta T(t)=(p_e(t)+\tau dp_e(t)/dt)/\alpha$. In other words, in deconvolution we search for the temperature the convolution of which is $p_e(t)$. This is a very useful result, which we are able to use as a new thermometer for global temperature. The theory is exact, but only noise or errors in the measured $\Delta T_m(t)$, $p_m(t)$, and $p_h(t)$ limit the accuracy of this method. Note that the method does not depend on the cause of the temperature change. The results of this work confirm our earlier studies about climate change \cite{Kau2011,Kau2014,Kau2018,Kau2019}. Since 1970, according to the observations, the changes of the low cloud cover have caused practically the observed temperature changes \cite{Kau2014,Kau2018,Kau2019,atmos12101297}. The low cloud cover has gradually decreased starting in 1975. The human contribution was about $0.01\degree$C in 1980 and now it is close $0.03\degree$C.

\bibliographystyle{unsrt}
\bibliography{Climate}
\end{document}